\documentclass{article} 
\usepackage{iclr2025_conference,times}

\usepackage{amsmath,amsfonts,bm}

\def\eqref#1{equation~\ref{#1}}

\def\1{\bm{1}}

\DeclareMathAlphabet{\mathsfit}{\encodingdefault}{\sfdefault}{m}{sl}
\SetMathAlphabet{\mathsfit}{bold}{\encodingdefault}{\sfdefault}{bx}{n}

\usepackage{hyperref}
\usepackage{url}
\usepackage{multirow}
\usepackage{rotating}
\usepackage{enumitem}

\usepackage{cleveref}
\title{Predicting Pulmonary Hypertension in \\Newborns: A Multi-view VAE Approach}

\author{Lucas Erlacher*, Samuel Ruipérez-Campillo*, \\
*Co-first Authors \\
Department of Computer Science\\
ETH Zurich\\
\texttt{elucas@ethz.ch, samuel.ruiperezcampillo@inf.ethz.ch} \\
\And
Holger Michel, Sven Wellmann, \\
Department of Neonatology, University Children’s Hospital Regensburg (KUNO)\\ Hospital St. Hedwig of the Order of St. John\\
University of Regensburg\\
\And
Thomas M. Sutter, Ece Ozkan$^\dagger$ \& Julia E. Vogt$^\dagger$ \\
$^\dagger$Co-senior Authors\\
Department of Computer Science\\
ETH Zurich\\
\texttt{\{thomas.sutter,  ece.oezkanelsen, julia.vogt\}@inf.ethz.ch} \\
}

\iclrfinalcopy 
\begin{document}

\maketitle

\begin{abstract}
Pulmonary hypertension (PH) in newborns is a critical condition characterized by elevated pressure in the pulmonary arteries, leading to right ventricular strain and heart failure. 
While right heart catheterization (RHC) is the diagnostic gold standard, echocardiography is preferred due to its non-invasive nature, safety, and accessibility. 
However, its accuracy highly depends on the operator, making PH assessment subjective. 
While automated detection methods have been explored, most models focus on adults and rely on single-view echocardiographic frames, limiting their performance in diagnosing PH in newborns. 
While multi-view echocardiography has shown promise in improving PH assessment, existing models struggle with generalizability.
In this work, we employ a multi-view variational autoencoder (VAE) for PH prediction using echocardiographic videos. 
By leveraging the VAE framework, our model captures complex latent representations, improving feature extraction and robustness. 
We compare its performance against single-view and supervised learning approaches. 
Our results show improved generalization and classification accuracy, highlighting the effectiveness of multi-view learning for robust PH assessment in newborns.
\end{abstract}

\section{Introduction}
Pulmonary hypertension (PH) in newborns is a progressive and potentially life-threatening condition characterized by elevated pulmonary artery pressure. 
It disrupts normal cardiopulmonary transition after birth, leading to impaired oxygenation, right ventricular overload, and high morbidity or mortality if left untreated \citep{Steinhorn2010, EL-Khuffash2014, Hansmann2017}.
Early and accurate diagnosis of neonatal PH is crucial for guiding therapeutic interventions.
However, diagnosing PH in newborns remains challenging due to small anatomical structures and dynamic cardiovascular changes that occur postnatally. 

Right heart catheterization (RHC) is the gold standard for diagnosing PH, providing precise hemodynamic measurements to assess the severity of the disease. 
However it is invasive, high-risk, and impractical for routine neonatal screening,  particularly due to its invasive nature and its risk for vascular and cardiac complications such as bleeding and infection \citep{Rosenkranz2015}. 
As a result, echocardiography has become the preferred alternative, offering a non-invasive, widely available, and real-time imaging modality for PH assessment in newborns.

Despite its advantages, echocardiographic screening for PH based on the visual evaluation of cardiac morphology and function—identifying abnormalities such as septal wall deviations, right ventricular hypertrophy, and changes in ventricular size—is highly subjective and operator-dependent \citep{Humbert2022}.
These qualitative assessments introduce inter-observer variability, making PH diagnosis inconsistent and potentially delaying life-saving interventions \citep{Augustine2018}.
Additional measurements in Doppler echocardiography or more recently introduced advanced techniques, such as Tissue-Doppler-derived methods, are time-consuming and require a very high level of expertise and training.
This highlights the urgent need for automated, objective, and scalable approaches for PH detection and severity prediction in newborns.

Recent advancements in machine learning have enabled automated echocardiographic image analysis for various cardiovascular conditions. 
However, most models are developed for adult PH patients, and their direct application to newborns is suboptimal due to anatomical and physiological differences. 
Additionally, many existing approaches rely on single-view echocardiographic frames, which may not provide comprehensive information for PH assessment.
In clinical practice, multi-view echocardiography is essential, as different echocardiographic views provide complementary information about cardiac function. 

To address these challenges, we adopt a multi-view variational autoencoder framework for PH prediction in newborns. 
VAEs excel at learning low-dimensional, structured latent representations from high-dimensional data, making them ideal for modeling echocardiographic videos. 
Given that multi-view learning conceptually align with multi-modal learning — where different data modalities (e.\,g.\,, images, text, and audio) are combined — we leverage multi-modal techniques to share information across different views. 
This allows us to utilize the strengths of multi-modal architectures while specifically adapting them to a multi-view echocardiographic setting, ensuring that information from different views is effectively utilized.

\paragraph{Our Contributions}
In this work, we propose an approach for PH prediction using a variational autoencoder (VAE) framework that leverages multi-view echocardiographic data. 
We hypothesize that extracting latent features from echocardiographic videos via a VAE-based representation will lead better generalization across different datasets, thereby enhancing downstream prediction. 
To this end, we introduce a two-stage pipeline that first encodes multi-view echocardiographic videos into a latent space, then uses these learned representations for PH classification. 
To unify the different views in a single framework, we incorporate a multi-view VAE with a variational mixture-of-experts prior (MMVM-VAE), aiming to capture both shared and view-specific features. 
In our experiments, we demonstrate that MMVM-VAE-based pre-training improves generalization compared to state-of-the-art supervised learning approaches. 

\section{Related Work}
To address the challenges of manual interpretation, machine learning (ML) techniques can improve PH detection using echocardiograms, particularly in newborns.
By analyzing echocardiographic datasets, ML models can identify subtle patterns associated with PH, reducing reliance on subjective assessments and enhancing diagnostic accuracy.
While numerous ML approaches have been developed for PH estimation in adults using imaging modalities such as chest X-rays, ECGs, CT scans, MRIs, and heart sounds \citep{Kaddoura2016, Dawes2017, Bello2019, Zou2020, Kusunose2020, Kwon2020, Mori2021, Vainio2021}, efforts focusing on the automated assessment of PH using echocardiography remain limited.

Two notable exceptions include \cite{Zhang2018}, which demonstrated the potential of deep learning for the prediction of PH using echocardiographic videos. 
However, their approach relies on static frames, failing to capture spatio-temporal patterns, and is limited to a single heart view. 
Later, \cite{Ragnarsdottir2024} introduced a multi-view and video-based deep learning approach using spatio-temporal CNNs, leveraging evidence that multiple views improve PH assessment accuracy \citep{Schneider2017}. 
Despite this improvement, their model exhibited limited generalizability on held-out test datasets, highlighting the need for further refinement and validation. 
Furthermore, most existing research has focused on adult patients, with relatively few studies dedicated to neonatal and pediatric populations \cite{Ragnarsdottir2024}. 
Methods that enhance generalizability are needed to better integrate different modalities and improve PH prediction performance in previously unseen newborn populations.

\section{Methods}
\label{sec:methods}

We consider a dataset $\mathbb{X} = \{ \bm{X}^{(i)} \}_{i=1}^n$ where each $\bm{X}^{(i)} = \{ \bm{x}_1^{(i)}, \ldots, \bm{x}_M^{(i)} \}$ is a set of $M$ different views $\bm{x}_m$ with corresponding latent variables $\bm{z} = \{ \bm{z}_1^{(i)}, \ldots, \bm{z}_M^{(i)} \}$.

\paragraph{Independent VAEs}
A variational autoencoder \citep[VAE, ][]{kingma2013} is a generative model that learns a probabilistic mapping between input data and a lower-dimensional latent space. 
In a single-view setting, each view $m$ is modeled independently with its own VAE, optimizing the following sum of evidence lower bounds (ELBOs) objective:
\begin{align}
\label{eq:independent_vaes}
     \mathcal{E} (\bm{X}) = &~ \sum_{m=1}^M \Big( \mathbb{E}_{q_\phi( \bm{z}_m \mid \bm{x}_m)} \left[ \log \frac{p_\theta (\bm{x}_m \mid \bm{z}_m)}{q_\phi( \bm{z}_m \mid \bm{x}_m)} \right] + \mathbb{E}_{q_\phi( \bm{z}_m \mid \bm{x}_m)} \left[ {\log p( \bm{z}_m )} \right] \Big).
\end{align}
Here, each view is encoded and decoded independently, where $p_\theta(\bm{x}_m \mid \bm{z}_m)$ is the decoder that reconstructs the input from the latent space, $q_\phi (\bm{z}_m \mid \bm{x}_m)$ is the encoder that approximates the posterior distribution over latent variables and, $\theta$ and $\phi$ are the learnable parameters of the respective models.

\paragraph{Multi-view VAE}
Optimizing \Cref{eq:independent_vaes}, we cannot leverage information shared between different views $\bm{X}_m$, leading to suboptimal generalization when integrating multiple echocardiographic views.
Previous works on multi-modal VAEs \citep{wu_multimodal_2018, shi_variational_2019, sutter2021} have demonstrated that incorporating relationships between data sources improves the quality of latent representations. 
These joint posterior approximation methods aggregate independent posteriors into a joint posterior but can suffer from poor reconstruction quality and, hence, inferior quality of representations. 
To overcome these limitations, we adopt the Multi-modal Variational Mixture Prior Model \citep[MMVM-VAE, ][]{sutter2024}, which uses a data-dependent prior, allowing the model to soft-share information between different echocardiographic views.

Unlike independent VAEs, the MMVM-VAE introduces a shared data-dependent prior distribution to encourage information exchange between different views and align their representations.
The prior ${h( \bm{z} \mid \bm{X} )}$, which models the dependency between the different views, is defined as
\begin{align}
    h(\bm{z} \mid \bm{X} ) = \prod_{m=1}^M h(\bm{z}_m \mid \bm{X}),~ \text{where} \hspace{0.25cm} h(\bm{z}_m \mid \bm{X}) = \frac{1}{M} \sum_{\tilde{m}=1}^M q_\phi (\bm{z}_m \mid \bm{x}_{\tilde{m}}), \hspace{0.25cm} \forall m \in M.
\vspace{-0.2cm}
\end{align}
This formulation ensures that the latent representation of each view is influenced by information from all available views instead of being learned independently. 

The MMVM-VAE optimizes the following objective that aligns independent posterior distributions: 
\begin{align}
    \mathcal{E} (\bm{X}) = &~ \sum_{m=1}^M \mathbb{E}_{q_\phi( \bm{z}_m \mid \bm{x}_m)} \left[ \log \frac{p_\theta (\bm{x}_m \mid \bm{z}_m)}{q_\phi( \bm{z}_m \mid \bm{x}_m)} \right] + \mathbb{E}_{q_\phi( \bm{z} \mid \bm{X})} \left[ {\log h( \bm{z} | \bm{X} )} \right].
\vspace{-0.2cm}
\end{align}
For more details, we refer to \citet{sutter2024}.

\section{Experiments and Results}

This section presents the experimental setup, dataset details, preprocessing pipeline, model architectures, evaluation metrics, results, and a discussion of our proposed approach for pulmonary hypertension classification from neonatal echocardiograms.

\subsection{Dataset}
The dataset used in this work is the same as in \cite{Ragnarsdottir2024}. 
It includes 936 transthoracic echocardiography videos collected from 192 newborns between 2019 and 2020,which are used for training and validation. 
Additionally, a held-out test set includes 375 videos from 78 newborns collected in 2022. 
All data were collected from a single medical center. 
A senior pediatric cardiologist performed the echocardiograms using a GE Logic S8 ultrasound machine with an S4-10 transducer operating at 6 MHz. 
Each video captures one of five standard heart views: PLAX (parasternal long axis), A4C (apical four-chamber), and three parasternal short-axis views—PSAX-P (papillary muscles), PSAX-S (semilunar valves), and PSAX-A (apical). 
Videos were recorded at 25 fps, with an average duration of 5 seconds, covering approximately 10 heartbeats.

\begin{figure}[t]
    \centering
    \includegraphics[width=\textwidth]{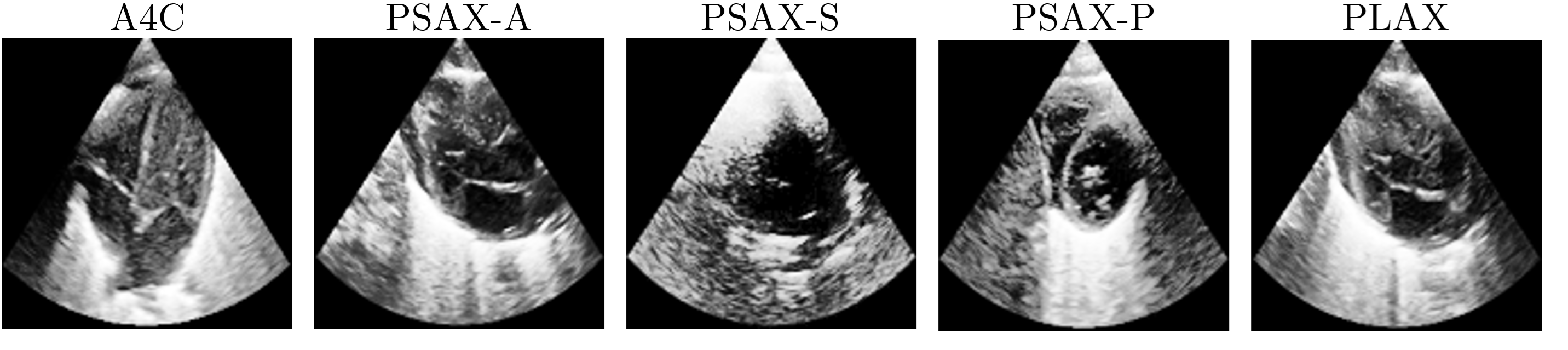}
    \caption{Example frames from the five standard echocardiographic views used in this study. Each row corresponds to a different view: PLAX (parasternal long axis), A4C (apical four-chamber), PSAX-P (parasternal short-axis at the papillary muscles), PSAX-S (parasternal short-axis at the semilunar valves), and PSAX-A (parasternal short-axis at the apex).}

    \label{fig:views}
\end{figure}

PH severity was labeled through expert assessment and categorized into three classes: none, mild, and moderate-to-severe. 
The dataset distribution varies between the two sets, with the training and validation set consisting of 65\% no PH, 17\% mild PH, and 18\% moderate-to-severe PH cases, while the test set contains 70\% no PH, 12\% mild PH, and 18\% moderate-to-severe PH. 
PH grading was determined based on the PSAX-P view, where no PH corresponds to the absence of septal flattening, mild PH is characterized by septal flattening into the right ventricle, and moderate-to-severe PH is identified by septal bowing into the left ventricle. Each video is also labeled by view. \Cref{fig:views} shows an example of the five views at a fixed frame for the same patient.

The study received ethical approval, and all data were pseudonymized to ensure privacy. 
For a more detailed overview, please refer to \cite{Ragnarsdottir2024}.

\subsection{Preprocessing and Data Augmentation}
Following \cite{Ragnarsdottir2024}, we preprocess the echocardiography videos by cropping and masking to remove extraneous information (e.g., annotations, additional signals) outside the scanning sector. 
The frames are resized to $128\times128$ pixels using bicubic interpolation. 
To normalize intensity distributions, we apply histogram equalization, which enhances contrast and ensures a more uniform distribution of pixel values across the grayscale range.

To enhance model robustness, we apply the following intensity and spatial transformations during training.
Intensity transformations improve invariance to brightness variations in ultrasound imaging. 
We randomly alter the brightness, contrast, saturation and hue values of an image in the interval $[0.3, 1.7]$, 
randomly apply a Gaussian blur with kernel size $(5, 5)$ and standard deviation $(3.5, 3.5)$ with a probability of $0.55$,
introduce salt-and-pepper noise with a threshold of $0.05$ and Gaussian noise with a standard deviation of $0.2$.
Spatial transformations increase robustness to variations in transducer positioning and zoom settings.
We employ random rotations between $-20^\circ$ and $20^\circ$ and translations up to 14\% of the image size, and
scale the image within a range of $\times$0.7 to $\times$1.2.

To address dataset imbalance, we employ a class balancing strategy using minority-class oversampling, ensuring a uniform class distribution in each training batch. 
This technique is applied only to the training sets to prevent the model from biasing towards the majority class, while the evaluation and test sets remain the same to preserve the integrity of performance metrics. 
Additionally, augmentation ensures each oversampled instance remains unique, reducing overfitting risk \cite{buda2018systematic}.

\subsection{Implementation Details}
\begin{figure}[t]
    \centering
    \includegraphics[width=\textwidth]{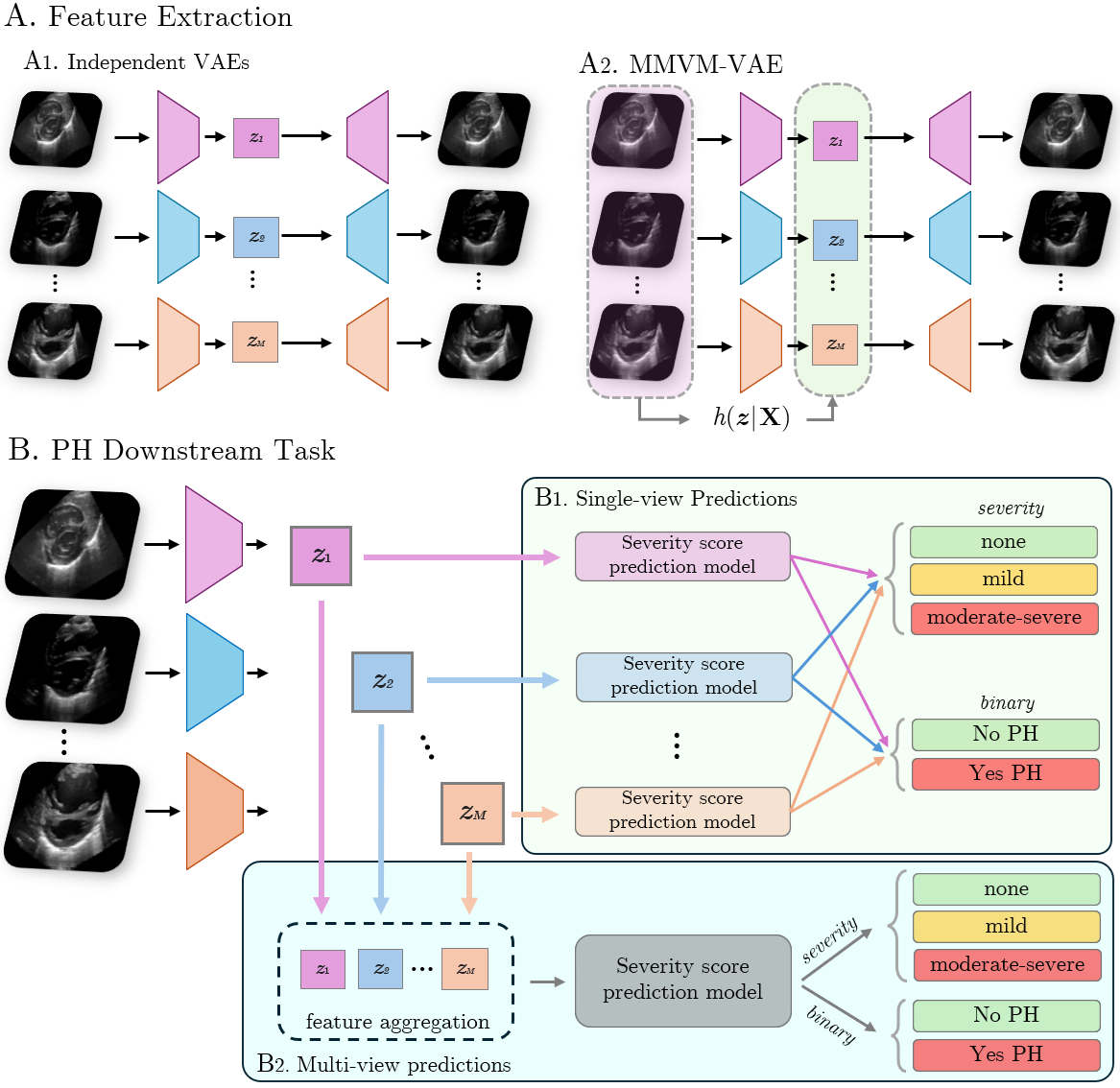}
    \caption{Schematic summary of the 2-step approach to predict PH from newborn echocardiograms. \textbf{A.} Illustration of the feature extraction mechanism on a VAE framework where individual VAEs are trained for each modality (A$_1$) or an MMVM-VAE is used to integrate common and single view information in the representation (A$_2$). \textbf{B.} After pretraining the encoders in A, the downstream task is performed by using one view at a time (single-view, B$_1$) or aggregating the latent representation of the different views (multi-view, B$_2$) to predict binary or severity PH.}
    \label{fig:arch}
\end{figure}

The methods used in this study are illustrated in \Cref{fig:arch}.

\paragraph{VAE Architecture}
We use multi-view VAEs to learn low-dimensional latent representations from multiple echocardiographic views. 
Each independent (single-view) encoder has four blocks, where each block includes three convolutional layers with padding (to preserve spatial dimensions), a batch normalization layer, a ReLU activation, and a max-pooling layer. 
The first convolution layer in each block doubles the number of channels. 
After the final block, outputs are flattened, then two fully connected layers produce the mean and exponentiated standard deviation for the latent space. 
Each independent (single-view) decoder is structured symmetrically to the encoder, comprising four blocks, where each block contains a transposed convolution (halves the channel dimension), two additional transposed convolutions, a batch normalization layer, and a ReLU activation. 
The latent vector is first passed through a fully connected layer, then reshaped to match the input format of the transposed convolution, ensuring a structured reconstruction. 

\paragraph{Classifier Architecture}
A three-layer feed-forward network classifies PH severity from the learned representations. Each fully connected layer is followed by batch normalization and ReLU. We use orthogonal weight initialization, and a \textit{softmax} at the output for class probabilities.

\paragraph{VAE Training}
We train the multi-view VAEs with an ELBO-based loss -- standard for individual VAEs \cite{kingma2013} and a variant for the MMVM-VAE approach \cite{sutter2024}-- to reconstruct ultrasound clips. Both PH-labeled and unlabeled samples are used at this stage.

\paragraph{Classifier Training}
We train the predictor on PH-labeled data using cross-entropy loss, while also fine-tuning the encoder weights. 
Although we experimented with various regularization methods (such as dropout and label smoothing), we found that a lightweight predictor network, combined with strong data augmentation, was sufficient to effectively prevent overfitting in our final models.

\subsection{Experiments}

\paragraph{Models} We evaluate three benchmark models for PH prediction. 
The supervised baseline is a deep learning model trained end-to-end using labeled echocardiographic videos, as proposed in \cite{Ragnarsdottir2024}. 
The second approach, independent VAEs, involves training separate VAEs for each echocardiographic view, followed by feature aggregation for downstream classification. 
Finally, we adopt MMVM-VAE, a multi-view variational autoencoder that learns a shared latent space across multiple views, incorporating both labeled and unlabeled data.

\paragraph{Views} We evaluate two approaches: a single-view approach, where predictions are made independently for each echocardiographic view, and a multi-view approach that integrates information across views. 
Within the multi-view approach, we consider three methods: (1) the supervised baseline, as in \cite{Ragnarsdottir2024}; (2) independent VAEs with feature aggregation, where latent representations from the VAEs are concatenated and used for classification; and (3) MMVM-VAE, where latent aggregation occurs within a shared latent space, facilitating more effective integration of multi-view information.

\paragraph{Downstream Tasks} We evaluate model performance on two classification tasks: binary PH detection, differentiating between newborns with and without pulmonary hypertension (PH); and PH severity prediction, categorizing PH severity into three levels: no PH, mild PH, and moderate-to-severe PH.

All models are trained using the same pre-processing pipeline, augmentation strategies, and class-balancing techniques. 
Performance is then evaluated on the held-out test set using three random seeds using AUROC, balanced accuracy, and F1-score as primary metrics.

\subsection{Results}

\begin{table}[t]
    \centering
    \resizebox{\textwidth}{!}{
    \begin{tabular}{cllcccccc}
        \hline \noalign{\vskip 1mm}
        \multirow{3}{*}{\textbf{Category}} & \multirow{3}{*}{\textbf{Method}} & \multirow{3}{*}{\textbf{View}} 
        & \multicolumn{3}{c}{\textbf{Binary PH Detection}} & \multicolumn{3}{c}{\textbf{PH Severity Prediction}} \vspace{0.5mm} \\
        \cline{4-9} \noalign{\vskip 1mm}
        & & & \multirow{2}{*}{AUROC} & \multirow{2}{*}{\shortstack{F1 Score}} & \multirow{2}{*}{\shortstack{Balanced Acc.}} 
        & \multirow{2}{*}{AUROC} & \multirow{2}{*}{\shortstack{F1 Score}} & \multirow{2}{*}{\shortstack{Balanced Acc.}} \\
        & & & & & & & & \\
        \hline \noalign{\vskip 1mm}
        \multirow{10}{*}{\shortstack{Single\\View}} 
            & \multirow{5}{*}{Ind-VAE} 
                & A4C      & 0.71 $\pm$ 0.04 & 0.71 $\pm$ 0.02 & 0.66 $\pm$ 0.03 & 0.70 $\pm$ 0.03 & 0.53 $\pm$ 0.07 & \textit{0.53 $\pm$ 0.03} \\
                & & PLAX   & 0.59 $\pm$ 0.03 & 0.66 $\pm$ 0.03 & 0.60 $\pm$ 0.04 & 0.66 $\pm$ 0.05 & 0.37 $\pm$ 0.20 & 0.48 $\pm$ 0.08 \\
                & & PSAX-P & 0.79 $\pm$ 0.01 & \textit{0.74 $\pm$ 0.03} & 0.72 $\pm$ 0.01 & 0.60 $\pm$ 0.06 & 0.32 $\pm$ 0.02 & 0.40 $\pm$ 0.06 \\
                & & PSAX-S & 0.63 $\pm$ 0.05 & 0.59 $\pm$ 0.03 & 0.59 $\pm$ 0.08 & 0.56 $\pm$ 0.03 & 0.30 $\pm$ 0.03 & 0.37 $\pm$ 0.04 \\
                & & PSAX-A & 0.70 $\pm$ 0.03 & 0.66 $\pm$ 0.09 & 0.67 $\pm$ 0.05 & 0.55 $\pm$ 0.03 & 0.34 $\pm$ 0.07 & 0.32 $\pm$ 0.04 \vspace{0.5mm} \\        
        \cline{2-9} \noalign{\vskip 1mm}
            & \multirow{5}{*}{\shortstack{MMVM-VAE}} 
                & A4C      & 0.58 $\pm$ 0.05 & 0.60 $\pm$ 0.03 & 0.52 $\pm$ 0.02 & 0.55 $\pm$ 0.03 & 0.56 $\pm$ 0.01 & 0.34 $\pm$ 0.01 \\
                & & PLAX   & 0.52 $\pm$ 0.01 & 0.57 $\pm$ 0.01 & 0.51 $\pm$ 0.01 & 0.54 $\pm$ 0.06 & 0.39 $\pm$ 0.25 & 0.37 $\pm$ 0.05 \\
                & & PSAX-P & 0.59 $\pm$ 0.12 & 0.61 $\pm$ 0.08 & 0.57 $\pm$ 0.10 & 0.53 $\pm$ 0.03 & 0.33 $\pm$ 0.23 & 0.34 $\pm$ 0.03 \\
                & & PSAX-S & 0.50 $\pm$ 0.00 & 0.56 $\pm$ 0.00 & 0.50 $\pm$ 0.00 & 0.58 $\pm$ 0.08 & 0.41 $\pm$ 0.28 & 0.37 $\pm$ 0.03 \\
                & & PSAX-A & 0.57 $\pm$ 0.10 & 0.62 $\pm$ 0.09 & 0.56 $\pm$ 0.09 & 0.59 $\pm$ 0.08 & 0.41 $\pm$ 0.27 & 0.40 $\pm$ 0.10 \vspace{0.5mm} \\
        \hline \noalign{\vskip 1mm}
        \multirow{3}{*}{\shortstack{Multi\\View}} 
            & Supervised & All views & \textbf{0.86 $\pm$ 0.08} & \textbf{0.80 $\pm$ 0.04} & \textbf{0.77 $\pm$ 0.04 }& 0.72 $\pm$ 0.05 & 0.58 $\pm$ 0.03 & 0.42 $\pm$ 0.03 \\
            & Ind-VAE-FA & All views & 0.78 $\pm$ 0.04 & 0.71 $\pm$ 0.02 & 0.69 $\pm$ 0.02 & \textit{0.73 $\pm$ 0.02} & \textbf{0.65 $\pm$ 0.02} & 0.50 $\pm$ 0.02 \\ 
            & MMVM-VAE (us) & All views &\textit{ 0.83 $\pm$ 0.02} & 0.72 $\pm$ 0.03 &\textit{ 0.74 $\pm$ 0.02} &\textbf{ 0.74 $\pm$ 0.01} & \textit{0.63 $\pm$ 0.03} & \textbf{0.56 $\pm$ 0.04} \vspace{0.5mm} \\
        \hline \hline
    \end{tabular}
    }
    \caption{Evaluation of binary PH detection and PH severity classification across different methods and echocardiographic views. Models are assessed using AUROC, F1-score, and balanced accuracy. Single-view models predict PH from individual echocardiographic views, while multi-view models integrate information across multiple views. Ind-VAE refers to independent variational autoencoders trained per view, Ind-VAE-FA and and MMVM-VAE aggregate feature representations before classification. The highest numbers in each column are represented in \textbf{bold} numbers, and second highest in \textit{italics.}}
    \label{tab:results_full_data}
\end{table}

\Cref{tab:results_full_data} summarizes our main findings for both binary PH detection and multi-class PH severity prediction under single-view and multi-view settings.

We observe that incorporating multiple standard echocardiographic views in our MMVM-VAE framework substantially improves classification performance compared to relying on a single-view input. 
Our multi-view MMVM-VAE model yields stable results, particularly for the multi-class PH severity prediction task, where it achieves higher balanced accuracy and a strong AUROC relative to alternative approaches (\Cref{tab:results_full_data}).

Importantly, the multi-view framework partially mitigates the drop in performance when moving from validation to unseen test data. 
While this effect can't directly be observed in \Cref{tab:results_full_data}, supplementary analyses in \Cref{tab:results_val,tab:results_test} offer detailed comparisons between validation and test sets, thereby illustrating this trend.
Generalizing to new data in the newborns population can be notoriously difficult due to smaller patient numbers, subtle anatomic variations, and differences in image acquisition. 
By enforcing shared latent structures across views, the MMVM-VAE learned representations that are more robust to these variations.

\subsection{Limitations and Future Work}

Despite these promising findings, several limitations remain. 
First, while our approach exploits multiple echocardiographic views, these videos were not acquired simultaneously, meaning that physiologic variations --such as differences in respiration or cardiac cycle phase-- may exist between views. 
This mismatch can introduce noise in cross-view alignment or reduce the actual amount of shared information. 
Future work could explore synchronized acquisitions or cycle-specific temporal alignment techniques to address this. 
Second, our approach assumes that most standard views are available for each patient. 
However, in real-world clinical settings, some views may be missing or of poor quality, which could impact model performance. 
Addressing this challenge will require missing-view imputation or partial-view inference strategies to ensure robustness when complete data are unavailable. Third, our data was collected from a single medical center, leaving open questions about demographic differences. 
Demographic differences in neonatal cohorts and variations in imaging protocols could affect model performance. 
A critical next step is to evaluate these methods on larger, multi-center cohorts to generalize across diverse patient populations.

In future work, we plan to extend the MMVM-VAE approach by incorporating explicit spatio-temporal modeling to better capture dynamic cardiac function. 
Additionally, we will explore whether the learned representations can transfer to other clinical tasks, such as predicting response to therapy or risk of adverse outcomes. 
Further, we aim to explore alternative multi-view latent aggregation strategies beyond the mixture-of-experts prior to further delineate how best to capture both shared and view-specific echocardiographic patterns. 

Overall, these results support the utility of multi-view VAE-based representations for improving PH screening and severity classification in newborns, advancing toward a more reliable and less operator-dependent diagnostic process.

\section{Conclusion}
Our results support the effectiveness of multi-view learning and VAE-based representations for improved PH screening and severity classification in newborns. By consolidating complementary information across standard echocardiographic views, our MMVM-VAE framework achieved more robust detection and grading of PH, with more consistency and less decreasing in performance when translating to a held-out test cohort, as compared to the supervised approach. 
This work represents a step toward a more personalized diagnosis and automated diagnostic process, reducing operator dependency and improving early and reliable PH evaluation in newborns. 


\bibliographystyle{iclr2025_conference}
\bibliography{ref.bib}

\newpage
\appendix
\section{Supplementary Results}
In this appendix, we present additional analyses and extended experimental outcomes that complement the main findings reported in the manuscript. These supplementary results provide model variants—ranging from single-view approaches to our proposed multi-view VAE framework—perform under varying conditions and metrics. Specifically, we include detailed per-view comparisons in a 5-fold cross-validation setting.

\begin{table}[h]
    \centering
    \resizebox{\textwidth}{!}{
    \begin{tabular}{cllcccccc}
        \hline \noalign{\vskip 1mm}
        \multirow{3}{*}{\textbf{Category}} & \multirow{3}{*}{\textbf{Method}} & \multirow{3}{*}{\textbf{View}} 
        & \multicolumn{3}{c}{\textbf{Binary PH Detection}} & \multicolumn{3}{c}{\textbf{PH Severity Prediction}} \vspace{0.5mm} \\
        \cline{4-9} \noalign{\vskip 1mm}
        & & & \multirow{2}{*}{AUROC} & \multirow{2}{*}{\shortstack{F1 Score}} & \multirow{2}{*}{\shortstack{Balanced Acc.}} 
        & \multirow{2}{*}{AUROC} & \multirow{2}{*}{\shortstack{F1 Score}} & \multirow{2}{*}{\shortstack{Balanced Acc.}} \\
        & & & & & & & & \\
        \hline \noalign{\vskip 1mm}
        \multirow{15}{*}{\shortstack{Single\\View}} 
            & \multirow{5}{*}{Supervised} 
                & A4C    & 0.87 $\pm$ 0.04 & 0.83 $\pm$ 0.04 & 0.83 $\pm$ 0.03 & 0.79 $\pm$ 0.04 & 0.75 $\pm$ 0.05 & 0.67 $\pm$ 0.06 \\
                & & PLAX   & 0.92 $\pm$ 0.05 & 0.88 $\pm$ 0.04 & 0.88 $\pm$ 0.04 & 0.84 $\pm$ 0.04 & 0.76 $\pm$ 0.04 & 0.70 $\pm$ 0.05 \\
                & & PSAX-P & 0.93 $\pm$ 0.04 & 0.91 $\pm$ 0.03 & 0.92 $\pm$ 0.03 & 0.83 $\pm$ 0.03 & 0.81 $\pm$ 0.02 & 0.74 $\pm$ 0.03 \\
                & & PSAX-S & 0.83 $\pm$ 0.03 & 0.81 $\pm$ 0.03 & 0.81 $\pm$ 0.03 & 0.74 $\pm$ 0.07 & 0.68 $\pm$ 0.06 & 0.62 $\pm$ 0.06 \\
                & & PSAX-A & 0.86 $\pm$ 0.04 & 0.85 $\pm$ 0.03 & 0.84 $\pm$ 0.03 & 0.80 $\pm$ 0.03 & 0.75 $\pm$ 0.04 & 0.66 $\pm$ 0.04 \vspace{0.5mm} \\        
        \cline{2-9} \noalign{\vskip 1mm}
            & \multirow{5}{*}{Ind-VAE} 
                & A4C    & 0.68 $\pm$ 0.16 & 0.64 $\pm$ 0.10 & 0.66 $\pm$ 0.12 & 0.55 $\pm$ 0.15 & 0.46 $\pm$ 0.13 & 0.45 $\pm$ 0.20 \\
                & & PLAX   & 0.70 $\pm$ 0.10 & 0.71 $\pm$ 0.10 & 0.73 $\pm$ 0.12 & 0.50 $\pm$ 0.09 & 0.32 $\pm$ 0.18 & 0.44 $\pm$ 0.10 \\
                & & PSAX-P & 0.72 $\pm$ 0.07 & 0.70 $\pm$ 0.06 & 0.74 $\pm$ 0.06 & 0.62 $\pm$ 0.11 & 0.58 $\pm$ 0.09 & 0.48 $\pm$ 0.15 \\
                & & PSAX-S & 0.60 $\pm$ 0.11 & 0.52 $\pm$ 0.14 & 0.60 $\pm$ 0.15 & 0.55 $\pm$ 0.08 & 0.27 $\pm$ 0.18 & 0.42 $\pm$ 0.16 \\
                & & PSAX-A & 0.49 $\pm$ 0.11 & 0.54 $\pm$ 0.13 & 0.52 $\pm$ 0.18 & 0.55 $\pm$ 0.11 & 0.39 $\pm$ 0.18 & 0.42 $\pm$ 0.17 \vspace{0.5mm} \\
        \cline{2-9} \noalign{\vskip 1mm}
            & \multirow{5}{*}{MMVM-VAE} 
                & A4C    & 0.63 $\pm$ 0.08 & 0.64 $\pm$ 0.13 & 0.63 $\pm$ 0.06 & 0.59 $\pm$ 0.15 & 0.47 $\pm$ 0.24 & 0.42 $\pm$ 0.09 \\
                & & PLAX   & 0.68 $\pm$ 0.09 & 0.70 $\pm$ 0.13 & 0.64 $\pm$ 0.10 & 0.54 $\pm$ 0.10 & 0.45 $\pm$ 0.23 & 0.39 $\pm$ 0.09 \\
                & & PSAX-P & 0.73 $\pm$ 0.11 & 0.78 $\pm$ 0.06 & 0.68 $\pm$ 0.5 & 0.54 $\pm$ 0.08 & 0.47 $\pm$ 0.23 & 0.40 $\pm$ 0.12 \\
                & & PSAX-S & 0.51 $\pm$ 0.04 & 0.64 $\pm$ 0.14 & 0.51 $\pm$ 0.04 & 0.52 $\pm$ 0.06 & 0.31 $\pm$ 0.26 & 0.36 $\pm$ 0.03 \\
                & & PSAX-A & 0.59 $\pm$ 0.13 & 0.51 $\pm$ 0.23 & 0.60 $\pm$ 0.07 & 0.57 $\pm$ 0.06 & 0.33 $\pm$ 0.20 & 0.35 $\pm$ 0.04 \vspace{0.5mm} \\
        \hline \noalign{\vskip 1mm}
        \multirow{3}{*}{\shortstack{Multi\\View}} 
            & Supervised & All views & 0.92 $\pm$ 0.02 & 0.90 $\pm$ 0.02 & 0.90 $\pm$ 0.01 & 0.84 $\pm$ 0.03 & 0.82 $\pm$ 0.03 & 0.73 $\pm$ 0.04 \\
            & Ind-VAE-FA & All views & 0.76 $\pm$ 0.14 & 0.72 $\pm$ 0.15 & 0.72 $\pm$ 0.01 & 0.70 $\pm$ 0.10 & 0.71 $\pm$ 0.13 & 0.56 $\pm$ 0.19 \\ 
            & MMVM-VAE & All views & 0.75 $\pm$ 0.05 & 0.70 $\pm$ 0.06 & 0.72 $\pm$ 0.08 & 0.69 $\pm$ 0.10 & 0.71 $\pm$ 0.12 & 0.56 $\pm$ 0.17 \vspace{0.5mm} \\
        \hline \hline
    \end{tabular}
    }
    \caption{Evaluation of binary PH detection and PH severity classification across different methods and echocardiographic views. Models are assessed using AUROC, F1-score, and balanced accuracy. Single-view models predict PH from individual echocardiographic views, while multi-view models integrate information across multiple views. Ind-VAE refers to independent variational autoencoders trained per view, Ind-VAE-FA aggregates feature representations before classification, and MMVM-VAE learns a shared latent space across views.}
    \label{tab:results_val}
\end{table}

\begin{table}[h]
    \centering
    \resizebox{\textwidth}{!}{
    \begin{tabular}{cllcccccc}
        \hline \noalign{\vskip 1mm}
        \multirow{3}{*}{\textbf{Category}} & \multirow{3}{*}{\textbf{Method}} & \multirow{3}{*}{\textbf{View}} 
        & \multicolumn{3}{c}{\textbf{Binary PH Detection}} & \multicolumn{3}{c}{\textbf{PH Severity Prediction}} \vspace{0.5mm}\\
        \cline{4-9} \noalign{\vskip 1mm}
        & & & \multirow{2}{*}{AUROC} & \multirow{2}{*}{\shortstack{F1 Score}} & \multirow{2}{*}{\shortstack{Balanced Acc.}} 
        & \multirow{2}{*}{AUROC} & \multirow{2}{*}{\shortstack{F1 Score}} & \multirow{2}{*}{\shortstack{Balanced Acc.}} \\
        & & & & & & & & \\
        \hline \noalign{\vskip 1mm}
        \multirow{15}{*}{\shortstack{Single\\View}} 
            & \multirow{5}{*}{Supervised} 
                & A4C      & 0.81 $\pm$ 0.06 & 0.71 $\pm$ 0.06 & 0.68 $\pm$ 0.07 & 0.70 $\pm$ 0.10 & 0.54 $\pm$ 0.04 & 0.39 $\pm$ 0.04 \\
                & & PLAX   & 0.79 $\pm$ 0.07 & 0.73 $\pm$ 0.05 & 0.66 $\pm$ 0.05 & 0.76 $\pm$ 0.06 & 0.64 $\pm$ 0.05 & 0.43 $\pm$ 0.08 \\
                & & PSAX-P & 0.90 $\pm$ 0.04 & 0.79 $\pm$ 0.04 & 0.77 $\pm$ 0.04 & 0.79 $\pm$ 0.02 & 0.60 $\pm$ 0.05 & 0.47 $\pm$ 0.05 \\
                & & PSAX-S & 0.90 $\pm$ 0.05 & 0.80 $\pm$ 0.04 & 0.78 $\pm$ 0.04 & 0.73 $\pm$ 0.07 & 0.60 $\pm$ 0.07 & 0.46 $\pm$ 0.07 \\
                & & PSAX-A & 0.89 $\pm$ 0.05 & 0.83 $\pm$ 0.04 & 0.80 $\pm$ 0.05 & 0.84 $\pm$ 0.04 & 0.66 $\pm$ 0.05 & 0.49 $\pm$ 0.09  \vspace{0.5mm} \\
        \cline{2-9} \noalign{\vskip 1mm}
            & \multirow{5}{*}{Ind-VAE} 
                & A4C      & 0.67 $\pm$ 0.03 & 0.69 $\pm$ 0.04 & 0.66 $\pm$ 0.03 & 0.59 $\pm$ 0.04 & 0.39 $\pm$ 0.06 & 0.44 $\pm$ 0.06 \\
                & & PLAX   & 0.64 $\pm$ 0.05 & 0.69 $\pm$ 0.02 & 0.63 $\pm$ 0.04 & 0.52 $\pm$ 0.04 & 0.32 $\pm$ 0.07 & 0.37 $\pm$ 0.05 \\
                & & PSAX-P & 0.75 $\pm$ 0.05 & 0.69 $\pm$ 0.06 & 0.70 $\pm$ 0.04 & 0.58 $\pm$ 0.04 & 0.44 $\pm$ 0.10 & 0.41 $\pm$ 0.05 \\
                & & PSAX-S & 0.69 $\pm$ 0.04 & 0.59 $\pm$ 0.07 & 0.64 $\pm$ 0.04 & 0.50 $\pm$ 0.04 & 0.31 $\pm$ 0.10 & 0.31 $\pm$ 0.04 \\
                & & PSAX-A & 0.67 $\pm$ 0.03 & 0.64 $\pm$ 0.03 & 0.63 $\pm$ 0.03 & 0.55 $\pm$ 0.05 & 0.36 $\pm$ 0.13 & 0.38 $\pm$ 0.05  \vspace{0.5mm} \\
        \cline{2-9} \noalign{\vskip 1mm}
            & \multirow{5}{*}{MMVM-VAE} 
                & A4C      & 0.59 $\pm$ 0.07 & 0.57 $\pm$ 0.01 & 0.56 $\pm$ 0.06 & 0.55 $\pm$ 0.04 & 0.46 $\pm$ 0.21 & 0.39 $\pm$ 0.07  \\
                & & PLAX   & 0.57 $\pm$ 0.08 & 0.60 $\pm$ 0.03 & 0.55 $\pm$ 0.05 & 0.54 $\pm$ 0.03 & 0.42 $\pm$ 0.18 & 0.36 $\pm$ 0.06 \\
                & & PSAX-P & 0.62 $\pm$ 0.06 & 0.66 $\pm$ 0.05 & 0.60 $\pm$ 0.05 & 0.55 $\pm$ 0.06 & 0.38 $\pm$ 0.18 & 0.33 $\pm$ 0.05 \\
                & & PSAX-S & 0.52 $\pm$ 0.04 & 0.58 $\pm$ 0.03 & 0.51 $\pm$ 0.02 & 0.53 $\pm$ 0.04 & 0.31 $\pm$ 0.24 & 0.33 $\pm$ 0.01 \\
                & & PSAX-A & 0.58 $\pm$ 0.08 & 0.46 $\pm$ 0.14 & 0.56 $\pm$ 0.06 & 0.55 $\pm$ 0.03 & 0.37 $\pm$ 0.14 & 0.37 $\pm$ 0.03  \vspace{0.5mm} \\
        \hline \noalign{\vskip 1mm}
        \multirow{3}{*}{\shortstack{Multi\\View}} 
            & Supervised & All views & 0.86 $\pm$ 0.08 & 0.80 $\pm$ 0.04 & 0.77 $\pm$ 0.04 & 0.72 $\pm$ 0.05 & 0.58 $\pm$ 0.03 & 0.42 $\pm$ 0.03 \\ 
            & Ind-VAE-FA & All views & 0.68 $\pm$ 0.07 & 0.71 $\pm$ 0.05 & 0.66 $\pm$ 0.05 & 0.65 $\pm$ 0.02 & 0.62 $\pm$ 0.03 & 0.46 $\pm$ 0.04 \\ 
            & MMVM-VAE & All views & 0.78 $\pm$ 0.05 & 0.75 $\pm$ 0.05 & 0.75 $\pm$ 0.04 & 0.64 $\pm$ 0.04 & 0.64 $\pm$ 0.04 & 0.46 $\pm$ 0.06  \vspace{0.5mm} \\
        \hline \hline
    \end{tabular}
    }
    \caption{Model performance on the held-out test set, evaluated using AUROC, F1-score, and balanced accuracy. Results compare single-view and multi-view classification approaches for both binary PH detection and severity prediction. Model variants follow the same definitions as in Table \ref{tab:results_val}.}
    \label{tab:results_test}
\end{table}

\end{document}